\documentclass[review]{elsarticle}
\usepackage{graphics}
\usepackage{graphicx}
\usepackage{float}%place images here with [H]
\usepackage{bm}
\usepackage[nomarkers,notablist,nofiglist]{endfloat}
\usepackage{lineno,hyperref}
\modulolinenumbers[5]
\usepackage{siunitx}         
\usepackage{soul}
\usepackage{amsmath}
\usepackage{geometry} % to change the page dimensions
\usepackage{color} 
\usepackage{outlines}
\journal{arXiv}

\begin{document}

\begin{frontmatter}
	
	\title{Non-destructive Depth-Resolved Characterization of Residual Strain Fields in High Electron Mobility Transistors using Differential Aperture X-ray Microscopy}
	%% Group authors per af{}filiation:
	
	%% or include af{}filiations in footnotes:
	\author[1]{Darren C. Pagan\corref{mycorrespondingauthor}}
	\ead{dcp5303@psu.edu}
	
	\author[2]{Md Rasel}
	\author[1]{Rachel E. Lim}
	\author[3]{Dina Sheyfer}
	\author[3]{Wenjun Liu}
	\author[2]{Aman Haque}

	\cortext[mycorrespondingauthor]{Corresponding author}

	\address[1]{Materials Science and Engineering, The Pennsylvania State University}
	\address[2]{Mechanical Engineering, The Pennsylvania State University}
	\address[3]{X-Ray Science Division, Argonne National Laboratory}

	\begin{abstract}
    Localized residual stress and elastic strain concentrations in microelectronic devices often affect the electronic performance, resistance to thermomechanical damage, and, likely, radiation tolerance. A primary challenge for characterization of these concentrations is that they exist over sub-$\mu$m length-scales, precluding their characterization by more traditional residual stress measurement techniques. Here we demonstrate the use of synchrotron X-ray -based differential aperture X-ray microscopy (DAXM) as a viable, non-destructive means to characterize these stress and strain concentrations in a depth-resolved manner. DAXM is used to map two-dimensional strain fields between source and drain in a gallium nitride (GaN) layer within high electron mobility transistors (HEMTs) with sub-$\mu$m spatial resolution. Strain fields at various positions in both pristine and irradiated HEMT specimens are presented in addition to a preliminary stress analysis to estimate the distribution of various stress components within the GaN layer. $\gamma$-irradiation is found to significantly reduce the lattice plane spacing in the GaN along the sample normal direction which is attributed to radiation damage in transistor components bonded to the GaN during irradiation.
	\end{abstract}

\end{frontmatter}

% \section*{Figure List}

% \begin{outline}[enumerate]
%     \1 General figures
%         \2 Flow chart
%     \1 Thermal modeling figures
%         \2 Figure showing middle and end of track with rectilinear and regular (interpolated) grids
%     \1 CTE figures 
%         \2 \colorbox{green}{Expansion vs temperature}
%     \1 X-ray simulation figures
%         \2 Figure showing how we can move the beam around and get different temperature profiles
%         \2 Figure comparing the temperature profiles of the different P-V combinations (both rings and integrated)
%     \1 GPR figures
%         \2 Probably should have a basic figure to show how GPR works? (the sine wave example? but that also doesn't feel like a figure that should be in this paper...) DCP: Not Necessary
%         \2 Figures for each of the temperature metrics prediction DCP: Which Metrics
            
%             \3 Figures showing how addition of different pieces of data changes prediction (e.g. only training on one P-V set vs 8 of them)
%             \3 Inclusion of more microstructures per P-V Set
%     \1 Experimental figures
%         \2 \colorbox{green}{Figure showing experimental setup}
%         \2 Figure showing example of experimental data (1 frame with rings, integrated data heatmap)
%         \2 Figure with predicted temperature metrics at different depths

% \end{outline}

\section{Introduction}

Residual stress and accompanying elastic strain have been critical design considerations for microelectronics with the emergence of 90 nm silicon technology~\cite{thompson2006uniaxial}. Furthermore, engineered uniform strain fields have since been exploited in device channels sustain Moore’s Law of scaling~\cite{wang2019strained}, with the most popular example being strain-engineered metal-oxide-semiconductor field-effect transistors (MOSFETs)~\cite{hoyt2002strained}. While such intentional strain fields have enhanced device performance, the role of localized, process-induced (e.g., selective epitaxial growth, dual stress liner, and shallow trench isolation) residual stresses and strains~\cite{van2006thermo,stuer2018characterisation,feng2018residual} have not been benign \cite{li2020study,bailey2007predicting}. In particular, microelectronic devices with highly localized stress fields are susceptible to charge traps or defects, even when the components materials are of high quality. In addition, these residual stress fields can contribute to thermomechanical degradation and failure during normal operation~\cite{perpina2016thermal,rodgers2004prediction}. However, fundamental connections between localized residual stress fields and device degradation have yet to be established due to the difficulty of mapping these fields at appropriate length-scales in comparison to other quantities of interest such as temperature. 

Beyond electronic performance and thermomechanical degradation, residual stress localization likely also affects the accumulation of radiation damage, and subsequently device performance, in microelectronics. Radiation damage in electronic devices can be categorized as either long term cumulative effects or single event effects (SEEs)~\cite{huang2019overview}, and the exact mechanism(s) of radiation interaction depends on the mass, charge, kinetic energy of the particles, radiation flux, and total exposure time. Cumulative effects from non-ionizing particles removing atoms from the lattice generate displacement damage that can adversely affect carrier mobility and increase charge trap density in semiconductors. In contrast, during SEE damage, charged particles can cause atomic displacement which can spontaneously ionize the semiconductor to create electron hole pairs. In addition, depending upon the local electric field, charge may diffuse to the nearest electrode to create transient spurious signal or soft errors. For example, a memory cell may flip its state due to the spurious signal~\cite{khachatrian2019effect}, giving rise to a glitch (or a burnout for extreme cases). Regions with localized residual stresses may be more susceptible to all of these damage modes because the material is already away from the equilibrium (zero-stress) configuration. Unfortunately, very few studies examine the role of residual stress concentrations~\cite{gaillardin2014comparative,mahatme2012impact,gaillardin2011investigations}. Therefore, there is an opportunity to connect the role of localized, preexisting residual stress in electronic devices with radiation susceptibility, and mapping localized residual stress fields before and after irradiation is a first step towards this goal.

The challenges for mapping the strain over a unit device (i.e., a transistor) include: (i) strain varies as a function of depth through the device, (ii) the length-scale for localization features of interest is often $\mu$m, far below the length-scale of traditional residual stress characterization techniques, and (iii) the characterization should be performed non-destructively as to not alter the stress field present through the creation of new free surfaces. As a result, most of the literature addresses global strain effects on electrical performance or reliability. Bending tests (i.e., cantilever~\cite{kang2003effect}, three-~\cite{chang2009changes}, or four-point \cite{liu2015effect}) are commonly used, along with wafer curvature \cite{liu2013influence} or thinning~\cite{azize2010effect}. These techniques can measure global uniaxial or biaxial strain, but cannot resolve the strain field over a transistor to capture localization. This is a motivation for this effort as localized stress effects cannot be estimated by homogenizing or averaging localized stress fields, reinforcing a need for measurement capabilities that map residual strain or stress at the single transistor level.  

An important microelectronic application in which residual stresses play an important role are high electron mobility transistors (HEMTs). Unlike conventional field effect transistors, HEMTs do not depend on doping to create the channel. Rather, they exploit the spontaneous piezoelectric polarization in a heterostructure to sustain a high-density two-dimensional electron gas~\cite{meneghini2021gan}. Gallium Nitride (GaN) has various properties making them ideal for utilization in these HEMTs. GaN is a wide bandgap, very high breakdown voltage material that can survive extreme temperatures and operating voltages, making GaN optimal for high power and high frequency applications. GaN also exhibits facile recrystallization upon ion impact and has higher threshold energy for atomic displacement~\cite{pearton2021radiation} contributing to radiation hardness, which is important for space and harsh environment applications. However, the technology has yet to reach predicted electronic performance~\cite{meneghini2021gan} or radiation hardness~\cite{fleetwood2022radiation}, highlighting unresolved challenges for adoption.

X-ray and neutron diffraction have long been the primary tools to non-destructively characterize residual stress distributions in crystalline materials \cite{noyan2013residual}. The critical component of all diffraction-based residual stress characterization techniques is that from measurable changes in the angle or energy of a diffraction event, the spacing of lattice planes are determined. From these lattice plane spacings, elastic strains and stresses can be inferred, as the stretching of the crystal lattice is directly related to the local elastic strain and stress state. Essentially the crystal lattice itself is used as ‘strain gauge’ with the orientation of the gauge being parallel to the normal of a set of lattice planes. In particular, the sin$^2$psi technique has been used extensively to characterize residual strains and stresses in thin films \cite{noyan1995residual} and surface stresses in larger engineering components \cite{noyan1985x}, while the more recently developed synchrotron-based energy dispersive diffraction technique is capable of measuring 3D distributions of lattice strain and stress at resolutions approaching 100 $\mu$m \cite{croft2005strain,tsakalakos2006measurement,mach2017validating,strantza2018coupled,strantza2021effect}. Analogous neutron-based measurements exist that operate on similar principles, but generally provide significantly increased penetration depth, but reduced spatial resolution \cite{allen1985neutron,hutchings2005introduction}. However, these diffraction methods are generally unable to probe residual strain fields at sufficient resolution for studying residual stress concentrations in microelectronic devices.

While traditional laboratory X-ray and neutron measurements have insufficient resolution for characterizing stress concentrations of interest, the past twenty years have seen a proliferation of synchrotron-based micro- and nanoscale diffraction techniques that can probe lattice state at length-scales commensurate with microelectronic devices. Specifically, differential aperture X-ray microscopy (DAXM) is capable of mapping lattice plane spacing (and associated lattice strain) at $\mu$m length-scales and below, in two and three dimensions \cite{liu2004three,yang2004differential,ice2005polychromatic,levine2006x,okoro2014nondestructive}. Here, we demonstrate how DAXM is capable of addressing the previously outlined challenges for measuring residual strains and stresses in microelectronic devices by characterizing subsurface elastic strain fields in the GaN layer within HEMTs. In this paper, bold symbols will be reserved for vectors and tensors, while superscripts will indicate coordinate systems in which quantities are expressed.

\section{Materials}

\begin{figure}[h]
	\centering
	\includegraphics[width=0.5\textwidth]{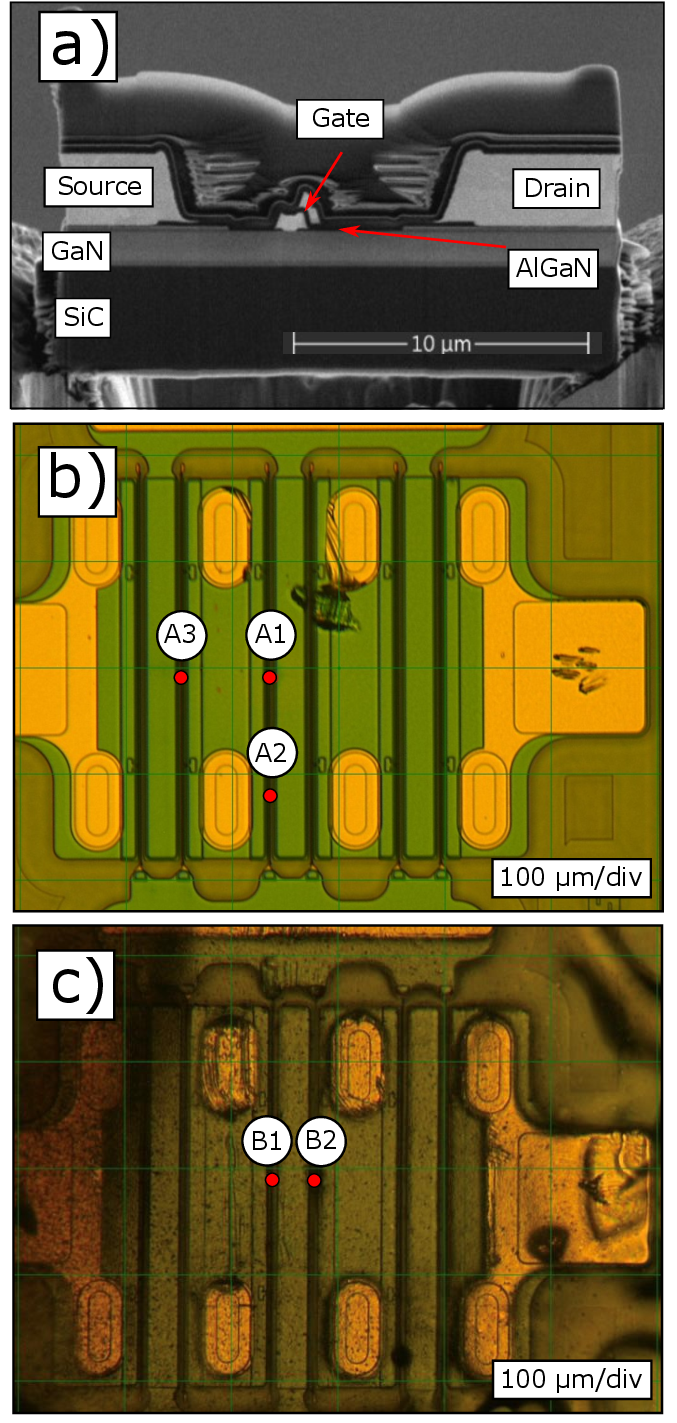}
	  \caption{a) Labeled cross section of a transistor channel in an example HEMT device. b) Optical image of the sample surface of the pristine HEMT specimen. The three positions at which DAXM measurements were performed are labeled A1, A2, and A3. c) Optical image of the sample surface of the irradiated HEMT specimen. The two positions at which DAXM measurements were performed are labeled B1 and B2.}
	  \label{fig:sample}
\end{figure}

Commercial HEMTs (CGH60008D, Wolfspeed) were used as specimens for the residual strain DAXM measurements. These commercial HEMTs are grown by MOCVD (metal-organic chemical vapor deposition) in a high-volume reactor on 100 mm semi-insulating 4H silicon carbide (4H-SiC) substrates. Ohmic contacts (e.g., Ti, Al, Ni, or Au) are formed directly on the top AlGaN layer, and Schottky metal gate electrodes are formed by recessing through a first SiN dielectric to the AlGaN and then depositing Ni, Pt, or Au metallization. To reshape and redistribute the very strong peak electric fields that occur at the drain-side edge, the gate electrode is laterally extended with gate metallization. Further electric field shaping is achieved by fabricating a source-connected second field plate after the second passivation layer \cite{pengelly2012review}. A cross section of the transistor configuration is given in Fig. \ref{fig:sample}a. The layer structure in the devices reported by the manufacturer include a 20 nm Al0.22Ga0.78N barrier, 1 nm thick AlN interlayer, 1.4 $\mu$m GaN buffer, and 100 $\mu$m 4H–SiC substrate with a gate length of 0.25 $\mu$m. The devices are approximately 600 $\mu$m $\times$ 600 $\mu$m in dimension and images of the devices can be viewed in Fig. \ref{fig:sample}.

Two devices were probed with residual strain measurements: one pristine device and one device that was irradiated. Irradiation was performed at room temperature to Co-60 $\gamma$ total ionizing dose (TID) of 5$\times 10^5$ rad (0.5 Mrad) at the Radiation Science and Engineering Center at Penn State University. The choice of TID for this work was informed by a previous study of $\gamma$ irradiation effects on GaN HEMTs for five different levels ranging from $10^5$ to $10^7$ rad \cite{rasel2022thermo}. In the previous study, appreciable damage from an electronic perspective (dislocation generation) was observed in transmission electron microscopy (TEM) measurements at $10^5$ rad. We also note that other sources of radiation (neutrons and charged particles) can generate significant radiation damage, but the non-activating nature of $\gamma$-irradiation was a benefit for technique development. Samples were fixed within a 4 in diameter $\times$ 4 in tall iso-dose region inside the $\gamma$-Cell and irradiated at a traceable certified TID rate of 180 krad/hr. Note that the damage or cloudy appearance on the specimens, particularly the irradiated specimen, is due to damage to an epoxy layer on the specimen surface, and is not necessarily indicative of damage to the device itself.

\section{Methods}

\begin{figure}[h]
	\centering
	\includegraphics[width=1.0\textwidth]{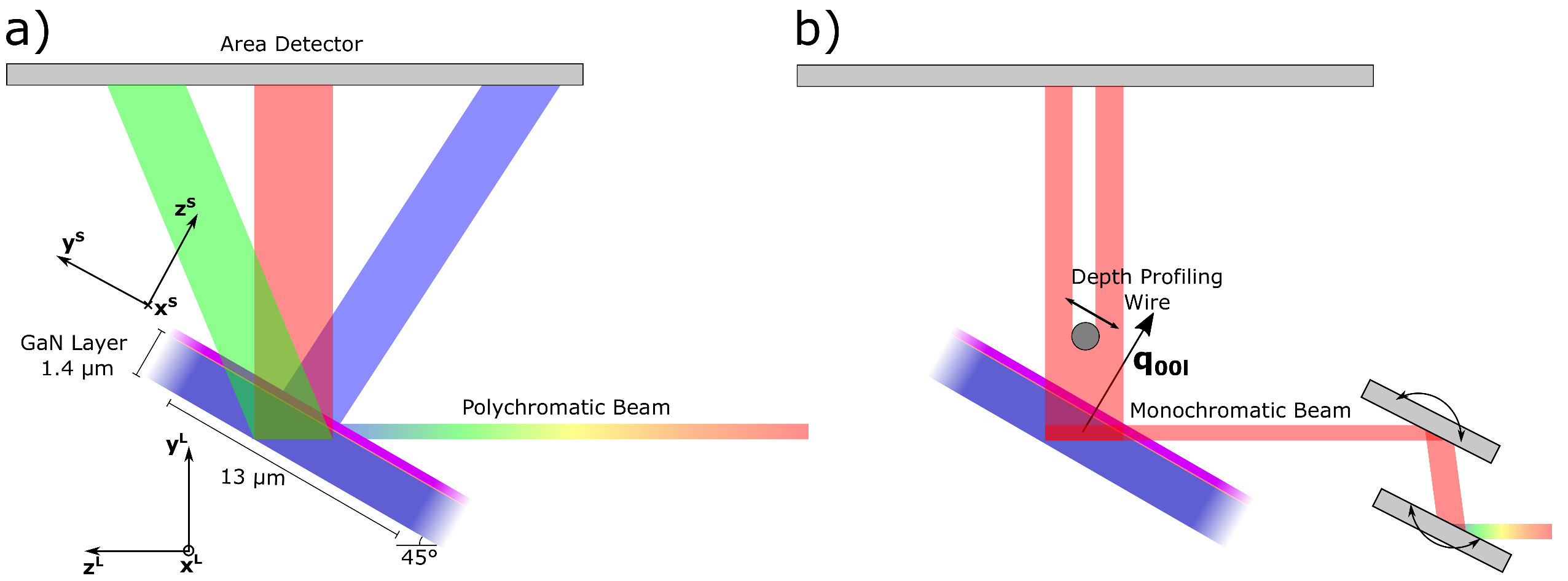}
	  \caption{a) Schematic of X-ray measurement configuration during indexing and orienting of the GaN layer. The specimen is illuminated by a polychromatic X-ray beam which produces multiple diffraction events. Each diffracted beam is emitted with a different energy (wavelength) from a different sets of lattice planes and captured on an area detector above the specimen. b) X-ray measurement configuration for DAXM measurements. The energy of the incoming monochromatic X-ray beam and position of platinum aperture wire are simultaneously varied to determine lattice plane spacing and position of diffraction events along the beam direction. Repeating the scans at different positions are then used to build 2D and 3D maps of lattice plane spacing.}
	  \label{fig:geom}
\end{figure}

Residual strain measurements within the GaN layer below individual transistors were performed at Beamline 34-ID-E at the Advanced Photon Source.  More complete descriptions of the instrument can be found in \cite{liu2014x}. An overview of the experiment geometry is given in Fig. \ref{fig:geom}a and Fig. \ref{fig:geom}b for polychromatic and monochromatic X-ray modes respectively. To aid discussion of measurements, we define a laboratory coordinate system denoted with L and a sample coordinate system with S. The incoming X-ray beam travels in the $\bm{z}^L$ direction and the vertical direction is $\bm{y}^L$ in the laboratory frame. The sample frame is defined with respect to the HEMT devices being probed with $\bm{x}^S$ and $\bm{y}^S$ aligned with the long sample edges and $\bm{z}^S$ being aligned with the device normal. For measurement, the X-ray beam was focused down to 400 nm along $\bm{x}^L$ and 700 nm along $\bm{y}^L$ using a set of Kirkpatrick–Baez (KB) mirrors. The sample was angled at 45$^\circ$ with respect to the incoming beam, so the focused X-ray had a footprint of 400 nm $\times$ 1000 nm on the samples. In polychromatic (white beam) mode, the X-ray spectrum nominally ranges from 7 to 30 keV. The X-ray beam is monochromated for diffraction peak energy measurements using a 2-bounce silicon monochromator that directs the beam to the same point on the specimen as the polychromatic beam. Lattice plane spacing is determined from Bragg’s law, reformulated to reflect that the energy of a diffraction event is being measured at a fixed Bragg angle $\theta$:
\begin{equation}
\lambda=\frac{hc}{E}=2d \sin(\theta)
\end{equation}
where $\lambda$ is the wavelength of the diffracting X-rays, $h$ is Plank’s constant, $c$ is the speed of light, $E$ is the energy of a diffracting X-ray, and $d$ is the spacing of a diffracting set of lattice planes. From the measured energy and lattice plane spacing, lattice strain $\varepsilon_{nn}$ is determined as:
\begin{equation}
\varepsilon_{nn} = \frac{d}{d_0}-1 = \bm{n} \cdot \bm{\varepsilon} \cdot \bm{n}
\label{eq:proj}
\end{equation}
where $\bm{n}$ is the normal to a set of lattice planes, $d_0$ is the unstrained spacing of a set of lattice planes, and $\bm{\varepsilon}$ is the full elastic strain tensor. From Eq. \ref{eq:proj} above, we interpret a lattice strain measurement as a projection of the elastic strain state along a lattice plane normal direction. 

Prior to strain measurements, various specimen alignments were completed. Rough positioning of the X-ray beam on the sample surface was performed using an optical microscope that was aligned to the beam position. Exactly locating individual transistors on the sample was performed by rastering the X-ray beam in 1 $\mu$m steps across source, gate, and drain on the sample surface while measuring X-ray fluorescence. Increased fluorescence from the gold within these features allowed for precision placement of the X-ray beam on a transistor. Determining the local orientation of the GaN layer with respect to the laboratory coordinate system was performed by illuminating the transistor using a polychromatic X-ray beam and collecting Laue diffraction peaks on a PerkinElmer amorphous Si area detector above the specimen (Fig. \ref{fig:geom}a). The detector was placed 0.5 m away from the specimen and calibration of the detector orientation and position was performed using diffraction peaks from a (001) silicon wafer. Note that when illuminated with a polychromatic beam, a single crystal will diffract X-ray beams in multiple directions, each with their own energy. Using the measured positions of GaN diffraction peaks on the area detector, the local orientation and, specifically, the lattice plane normal $\bm{n}$ of (00l) lattice planes (c-axis) was determined. In the HEMT devices, the c-axis was nominally aligned with normal to the device surface ($\bm{z}^S$).

After specimen alignment, depth-resolved measurement of lattice plane spacing in the GaN buffer layer between source and drain at various positions  on the two devices were performed (see Figs. \ref{fig:sample}b and \ref{fig:sample}c). As part of this process, a double bounce silicon monochromator was placed into the path of the incoming X-ray beam and a platinum wire (100 $\mu$m diameter) was placed above the specimen (Fig. \ref{fig:geom}b). Using the adjustable monochromator, the energy (wavelength) at which specific diffraction events occur is determined. From this information, the lattice plane spacing $d$ of the lattice planes participating in a diffraction event are determined. As the platinum wire is scanned through the diffracted X-ray beam, the position of diffraction events along the incoming beam direction (depth through the specimen) is also determined. Diffraction images were collected while: (i) translating the specimen to different position along $\bm{x}^S$ and then at each position (ii) scanning the wire (500 nm step size) through the diffracted beam to a create a virtual aperture through image subtraction at (iii) a series of monochromator angles (incident X-ray energies). From this series of diffraction images varying sample position, wire position, and monochromator angle, two-dimensional maps of lattice plane spacing of (008) sets of lattice planes were reconstructed using the LaueGo software package. 

The (008) lattice plane spacing were then scaled by a factor of 8 to determine the spacing of (001) lattice planes, equal to the c-axis lattice parameter. Lattice strains $\varepsilon_{nn}$ were then calculated using Eq. \ref{eq:proj} and a GaN c-axis lattice parameter from the literature. For lattice strain calculations, a nominal (001) lattice plane spacing of 5.1860~\AA \textvisiblespace was used for $d_0$. We note that there is a significant spread of reported lattice parameters in the literature generally ranging from 5.1855~\AA \textvisiblespace to 5.1865~\AA \textvisiblespace \cite{leszczynski1994thermal,leszczynski1996lattice,leszczynski1995lattice}. The choice of lattice parameter will scale lattice strains and even affect whether strains are interpreted as tensile or compressive. For this reason, lattice plane spacing along the c-axis direction and lattice strains are reported in tandem. Uncertainty for lattice strain measurements using DAXM has generally been determined to be less than $10^{-4}$.

\section{Results}

In this section, fields of (008) lattice plane spacing (and related lattice strains) within the GaN layer beneath the source, gate, and drain at various positions along both the pristine and irradiated HEMT devices are presented. Prior to presenting the fields, a general description of the measurements is provided. Each diffraction measurement (‘point’) corresponds to the average lattice plane spacing of a diffraction volume defined by the dimensions of the incoming X-ray beam (400 nm by 700 nm) and the projection of the differential aperture along the diffracting beam direction (500 nm). The spacing of the diffraction measurements were 1 $\mu$m along $\bm{x}^L$ and 350 nm along $\bm{z}^L$. Along $\bm{z}^L$, the diffraction measurements had a small amount of oversampling with approximately 75 nm of overlap between volumes. To aid the interpretation of the strain fields, the accompanying fluorescence measurements used to align the lattice plane spacing measurements are also provided. The increase of fluorescence at the center of each scan corresponds to the X-ray beam passing over the transistor gate. Sloping of the fluorescence scans is a shadowing effect caused by the fluorescence detector placement with respect to the incoming beam and sample. Lastly, we note that the arrangement of source, gate, and drain reverses depending on the position within the specimen. To facilitate comparison of the lattice plane spacing and strain fields at various positions on the HEMT device, measurements at A3 and B2 have been ‘mirrored’.

\begin{figure}[h]
	\centering
	\includegraphics[width=0.6\textwidth]{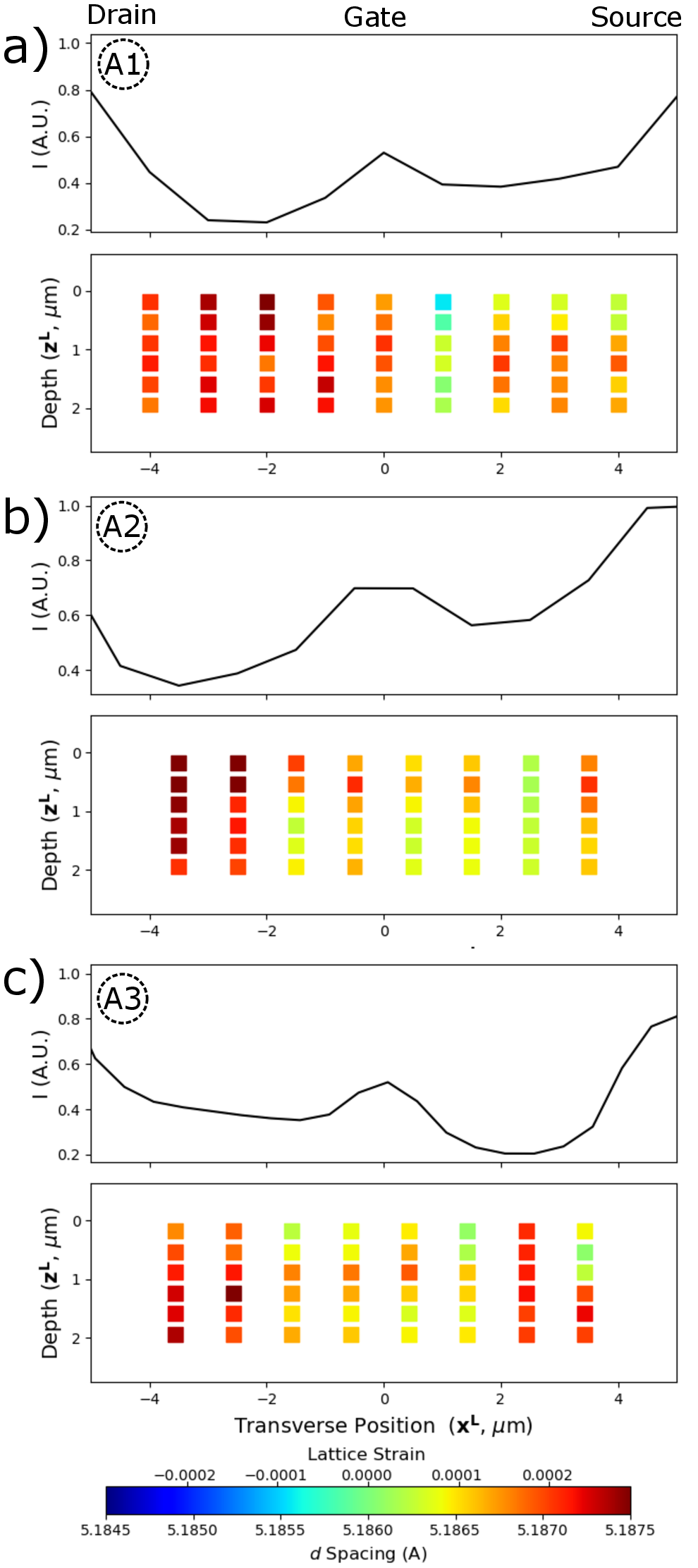}
	  \caption{Two-dimensional maps along of lattice plane spacing and strain of (001) lattice planes (along the c-axis) in the $\bm{x}^L$-$\bm{z}^L$ plane at points a) A1, b) A2, and c) A3 in the pristine HEMT sample. Above each two-dimension is the fluorescence intensity used for determining position between source and drain.}
	  \label{fig:pris}
\end{figure}

\begin{figure}[h]
	\centering
	\includegraphics[width=0.6\textwidth]{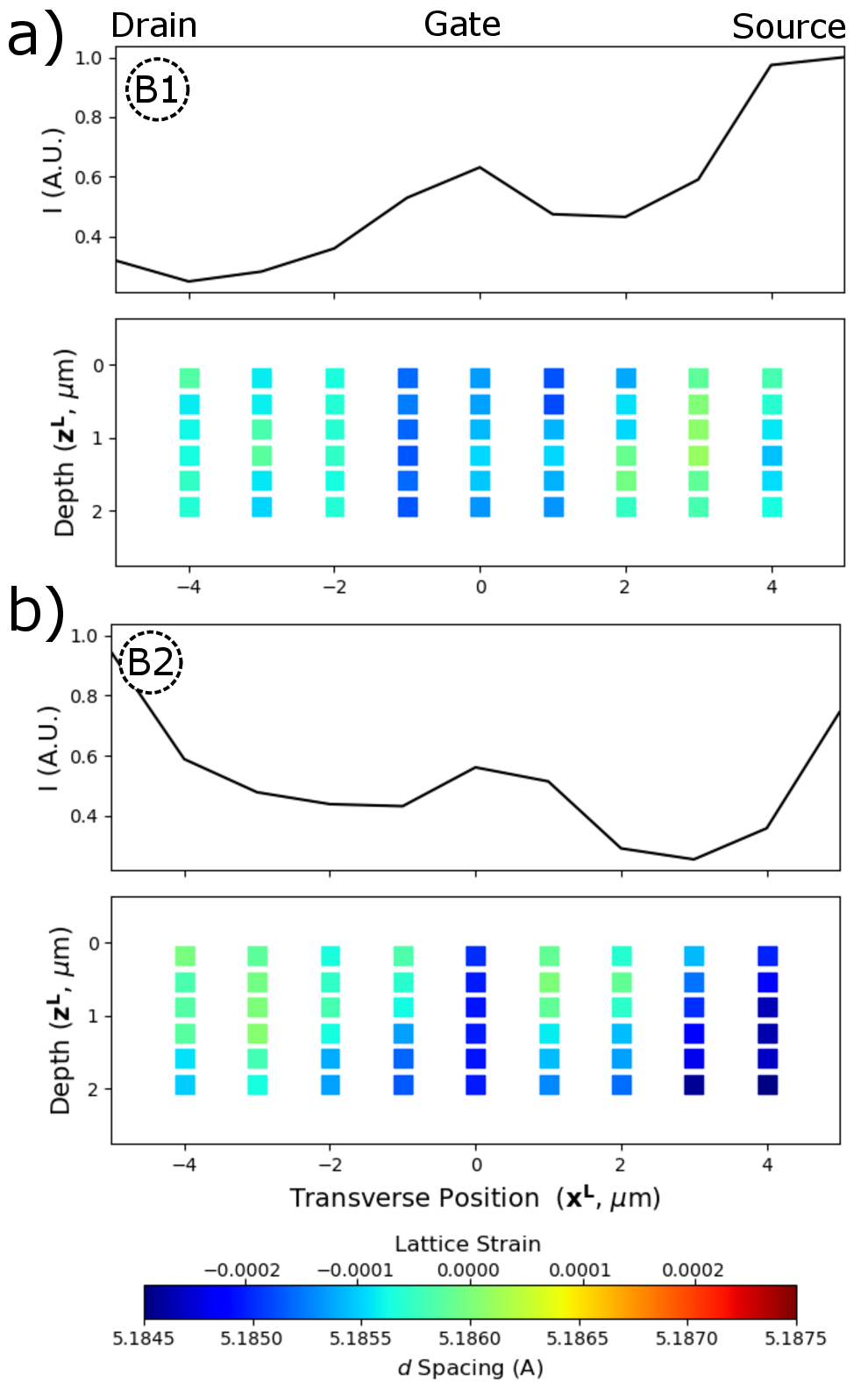}
	  \caption{Figure 4: Two-dimensional maps along of lattice plane spacing and strain of (001) lattice planes (along the c-axis) in the $\bm{x}^L$-$\bm{z}^L$ plane at points a) B1 and b) B2 in the irradiated HEMT sample. Above each two-dimension is the fluorescence intensity used for determining position between source and drain.}
	  \label{fig:irrad}
\end{figure}

Figure~\ref{fig:pris} shows the reconstructed distributions of lattice plane spacing and strain between source and drain at positions A1, A2, and A3 in the pristine HEMT specimen. With respect to the reference lattice plane spacing (5.186~\AA), the (001) lattice plane spacings are generally found to be greater than $d_0$ at the positions probed. The spread in lattice plane spacing across the measurement points is 2$\times10^{-3}$~\AA \textvisiblespace or 3$ \times 10^{-4}$ strain. In the figure, we can also see that across the positions on the pristine HEMT specimen, lattice plane spacings and strains are generally largest below the transistor drain and lowest below the gate. Similar to Fig.~\ref{fig:pris}, Fig.~\ref{fig:irrad} shows the reconstructed distributions of lattice plane spacing and strain between source and drain at points B1 and B2 in the irradiated HEMT specimen. The spread in lattice plane spacing and strains in the irradiated specimen is comparable, but notably the magnitude of lattice plane spacing is significantly decreased in the irradiated specimen. So much so, that the strains are now negative with the reference lattice place spacing used. Lastly, the largest strain magnitudes now appear to be below the gate as opposed to below the drain.

\section{Discussion}

As GaN-based HEMTs contain multilayer materials comprising the substrate, nucleation layer, GaN buffer layer, spacer, AlGaN layer, cap layer, and passivation layer, inherent residual stress and strain are present due to device fabrication. Fabrication of GaN-based devices requires multiple steps of depositing and annealing to ensure proper adhesion of layer stacks and metallic contacts.  As a result, pristine devices, even at ambient temperature and unbiased conditions, contain significant intrinsic or residual strain levels due to lattice mismatch and contraction of layers when the device cools from deposition to room temperatures. Unintentional or intentional doping of impurities/point defects can also create an additional strain known as hydrostatic strain~\cite{kisielowski1996strain}. Developing methods to characterize and understand these stresses and strains is critical, as stress and strain alter a device’s electron mobility, and sheet carrier density~\cite{zhang2000charge} and excessive tensile strain in the AlGaN layer under high voltage operations can result in electrically active defects and mechanical damage, resulting in performance deterioration and shortened lifetimes~\cite{del2009gan}. Furthermore, regions of existing high residual stress due to manufacture may make devices more susceptible to radiation damage and low TID of $\gamma$-irradiation is known to increase electrical output due to strain relaxation existing in mismatched heterostructures and structural ordering of native defects~\cite{lee2017low,kurakin2008mechanism,soltani2014high}. In this work, we demonstrated that the DAXM method provides an ideal means to characterize these residual strain and stress distributions influencing electronic performance due the spatial and strain resolutions achieved. In addition, the non-destructive nature of the measurements can open the path to \emph{in situ} measurements in the future.

\subsection{Stress Analysis}

\begin{figure}[h]
	\centering
	\includegraphics[width=1.0\textwidth]{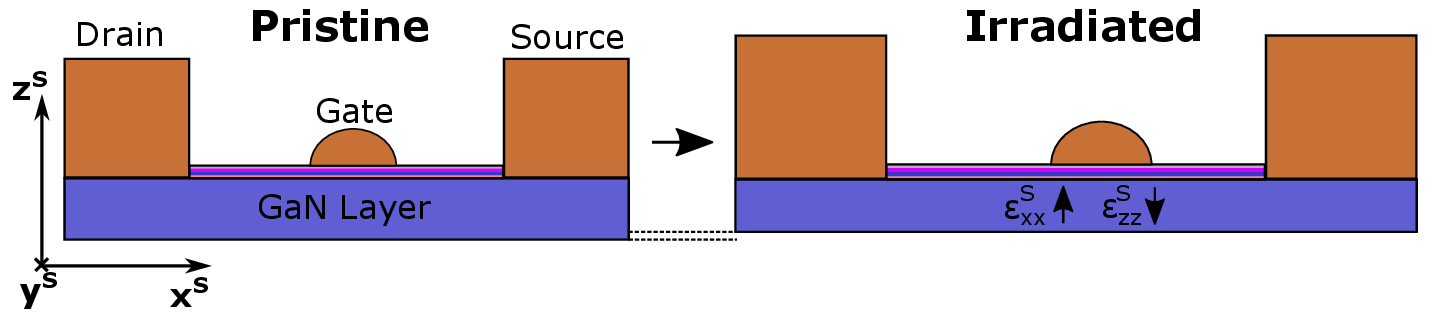}
	  \caption{Schematic of increased relative expansion of the adhered gold layers during irradiation driving a decrease in measured lattice strains perpendicular to the sample surface in the GaN layer.}
	  \label{fig:expan}
\end{figure}

As mentioned, the most striking difference between the strain fields in the pristine and irradiated samples is the significant reduction of the measured lattice plane spacing along the c-axis after irradiation. We note that similar results of decreasing c-axis lattice parameter after $6\times10^5$ rad $\gamma$ TID have been reported by Sharma et al. \cite{sharma2019cumulative}. In addition, the location of the largest strain magnitude and sign of stress both shift. Critical for the interpretation of the lattice strains reported is recognizing the lattice strains along the c-axis in the GaN layer ($\varepsilon_{zz}^S$ in the sample coordinate system) primarily reflect Poisson contraction or expansion due to transverse strains imposed by the deposited gold layers that comprise the source, gate and drain. Therefore, the decrease in lattice plane spacing between the pristine and irradiated specimens can be rationalized as increased expansion in the gold layers above the GaN due to irradiation causing increased tension along $\bm{x}^S$ and contraction along $\bm{z}^S$ (again the c-axis of the GaN is parallel to $\bm{z}^S$). This process is illustrated in Fig.~\ref{fig:expan} and is likely driven by an increase in point defect concentration after $\gamma$-irradiation in the GaN-based HEMT~\cite{polyakov2013radiation}. When $\gamma$-Rays interact with matter, they primarily interact with electrons and may get absorbed, scattered, or produce electron-positron pairs~\cite{nelson1991gamma}. More specifically, when $\gamma$-Rays interact with weakly bound electrons, they lose a portion of their energy to free the electron (Compton electrons) and are then scattered, which in turn can create Frenkel pairs and defect clusters which can migrate, recombine, or form complexes, changing the strain state within the material~\cite{kim2004comparison,hwang2014effect}.

Beyond the graphical interpretation of the deformation in the GaN layer in Fig.~\ref{fig:expan}, a relatively simple analysis can be performed to estimate the stress $\bm{\sigma}$ in the GaN layer. For simplicity, calculations will be performed in the sample coordinate system. For analysis, we employ the anisotropic form of Hookes’ Law (matrix Voigt notation) with hexagonal symmetry:
\begin{equation}
\begin{bmatrix}
C_{11} & C_{12} & C_{13} & 0 & 0 & 0 \\
C_{12} & C_{11} & C_{13} & 0 & 0 & 0 \\
C_{13} & C_{13} & C_{33} & 0 & 0 & 0 \\
0 & 0 &0 & C_{44} & 0 & 0 \\
0 & 0 &0 & 0  &C_{44} & 0 \\
0 & 0 &0 & 0 & 0 & 1/2(C_{11}-C_{12})
\end{bmatrix}
\begin{bmatrix}
\varepsilon_{xx}^S\\
\varepsilon_{yy}^S\\
\varepsilon_{zz}^S\\
2\varepsilon_{yz}^S\\
2\varepsilon_{xz}^S\\
2\varepsilon_{xy}^S
\end{bmatrix}
=
\begin{bmatrix}
\sigma_{xx}^S\\
\sigma_{yy}^S\\
\sigma_{zz}^S\\
\sigma_{yz}^S\\
\sigma_{xz}^S\\
\sigma_{xy}^S
\end{bmatrix}.
\label{eq:hooke}
\end{equation}
We note that as the c-axis is aligned with $\varepsilon_{zz}^S$ and GaN is transversely isotropic in the $\bm{x}^S$-$\bm{y}^S$ plane, Eq. \ref{eq:hooke} is valid for all orientations of the a-axes in the $\bm{x}^S$-$\bm{y}^S$ plane. In the GaN layer, we assume a plane strain state along the wires ($\varepsilon_{yy}^S$=0, see Fig.~\ref{fig:expan}) and a biaxial stress state ($\sigma_{zz}^S$=$\sigma_{yz}^S$=$\sigma_{xz}^S$=0) as a majority of it has a free surface within the transistor. Setting, the appropriate stress and strain components to 0 and reorganizing terms, we find that for the missing stress and strain components:
\begin{equation}
\begin{bmatrix}
1 & 0 & -C_{11} \\
0 & 1 & -C_{12}  \\
0 & 0 & -C_{13} 
\end{bmatrix}
\begin{bmatrix}
\sigma_{xx}^S\\
\sigma_{yy}^S\\
\varepsilon_{xx}^S
\end{bmatrix}
=
\begin{bmatrix}
C_{13} \varepsilon_{zz}^S\\
C_{13} \varepsilon_{zz}^S\\
C_{33} \varepsilon_{zz}^S
\end{bmatrix}
\label{fig:reorg}
\end{equation}
and
\begin{equation}
    \sigma_{xy}^S=2 C_{44} \varepsilon_{xy}.
\end{equation}
While the non-zero shear strain and shear stress components cannot be solved with the measured strain values (and  $\sigma_{xy}^S$ must be non-zero to maintain equilibrium), $\sigma_{xx}^S$,  $\sigma_{xx}^S$, and  $\varepsilon_{xx}^S$ are readily solved for using Eq. \ref{fig:reorg} and the single crystal elastic moduli of GaN. The moduli used for calculation are $C_{11}$=390 GPa, $C_{12}$=145 GPa, $C_{13}$=106 GPa, and $C_{33}$=398 GPa~\cite{polian1996elastic}.

\begin{figure}[h]
	\centering
	\includegraphics[width=1.0\textwidth]{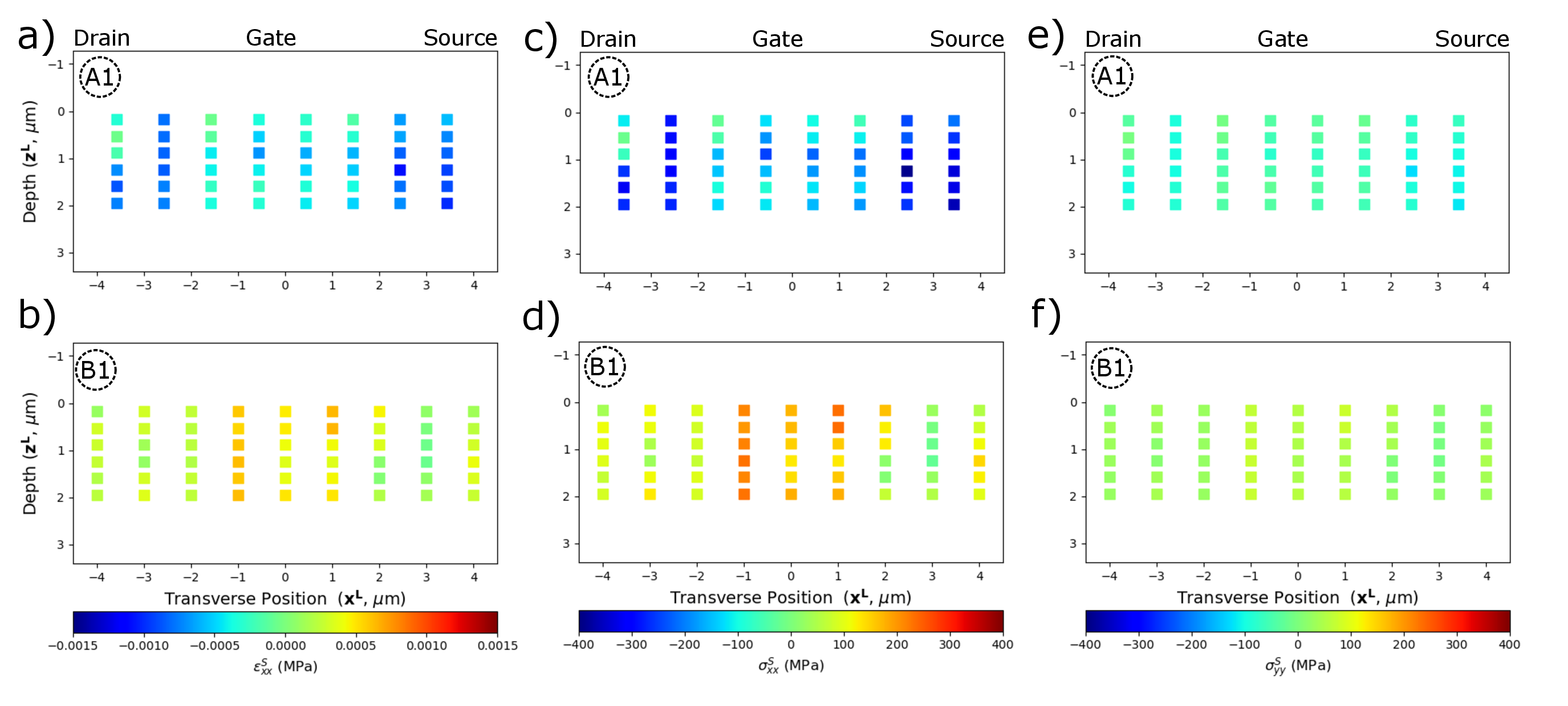}
	  \caption{Calculated strain component $\varepsilon_{xx}^S$ at a) position A1 in the pristine sample and b) position B2 in the irradiated specimen; calculated stress component $\sigma_{xx}^S$ at c) position A1 in the pristine sample and d) position B2 in the irradiated specimen; and calculated stress component $\sigma_{yy}^S$ at e) position A1 in the pristine sample and f) position B2 in the irradiated specimen.}
	  \label{fig:stress}
\end{figure}

Fig.~\ref{fig:stress}a, \ref{fig:stress}c, and \ref{fig:stress}e show the calculated distributions of $\varepsilon_{xx}^S$, $\sigma_{xx}^S$, and $\sigma_{yy}^S$ respectively at position A1 in the pristine sample, while the same for position B1 in the irradiated specimen are shown Fig.~\ref{fig:stress}b, \ref{fig:stress}d, and \ref{fig:stress}f. Since the trends are similar in the various measurement positions on each specimen, only one position per specimen is shown. As can be seen between the two specimens, the calculated stresses and strains generally follow the schematic in Fig. \ref{fig:expan}. In the pristine sample, the GaN is primarily under compression along the $\bm{x}^S$ direction ($\sigma_{xx}^S$) and with increased irradiation, the GaN is pulled into tension along $\bm{x}^S$ ($\sigma_{xx}^S$) by the expanding gold layers. In both specimens, the $\sigma_{yy}^S$ component is negligible. Importantly, with irradiation, the largest stress concentrations transition from underneath the drain to underneath the gate. We can compare these stress fields to others found in the literature. Other works have found the largest gradient of stress in pristine samples to be near the metal electrodes in GaN-based devices which are attributed to the stress present in the ohmic/Schottky gate electrodes \cite{choi2013analysis,matsuki2007impact}. Choi et al. have shown that the highest tensile stress is present near the edge of the metallization structure by photoluminescence spectroscopy (PL) and the bulk average value measured by Raman spectroscopy remains almost constant across the channel. With regards to the irradiated specimen, a recent study of $10^7$  Mrad $\gamma$-irradiation on a GaN-based HEMT mapped the spatial strain distribution along the channel before and after irradiation with nanobeam electron diffraction~\cite{rasel2022thermo}. A higher strain gradient near the device terminal after irradiation was observed and it was hypothesized that the highly localized stress regions were more susceptible to radiation-induced degradation.

\begin{figure}[h]
	\centering
	\includegraphics[width=0.6\textwidth]{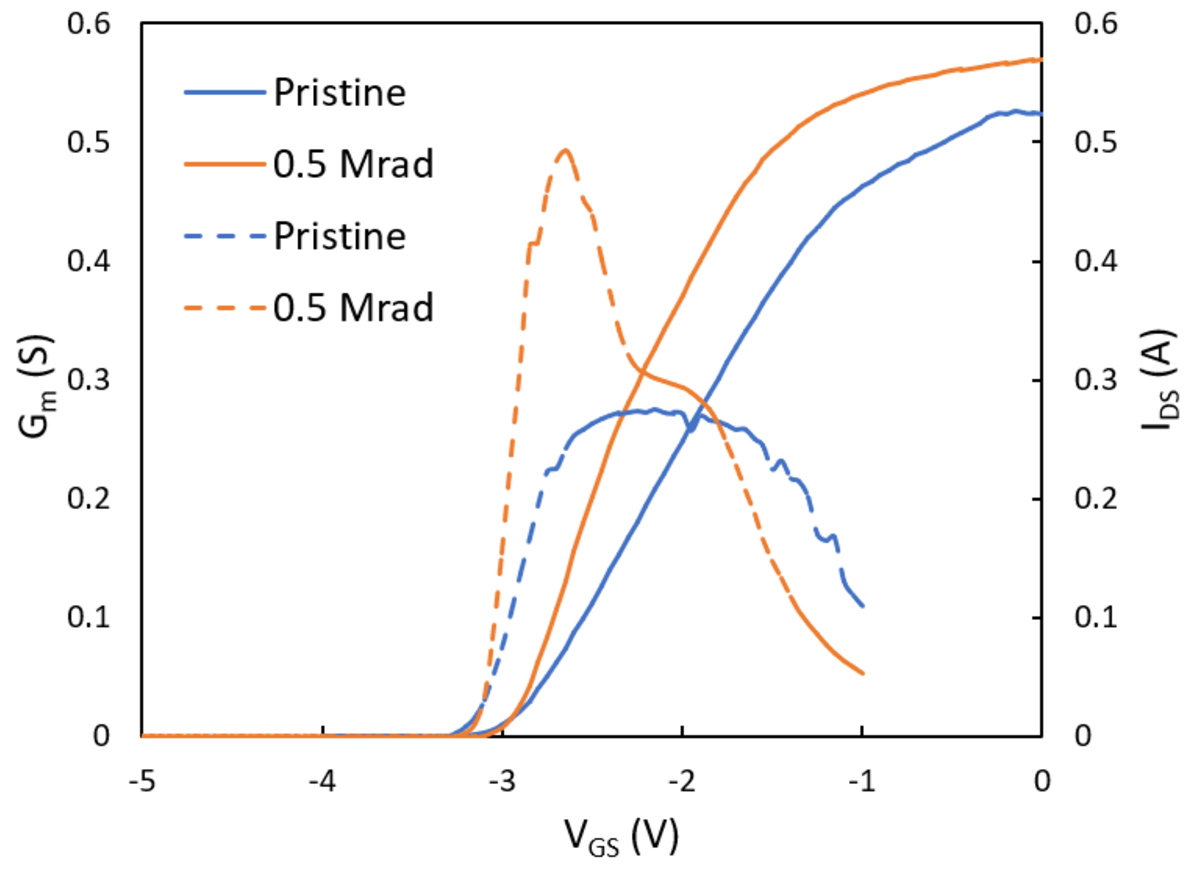}
	  \caption{Transfer characteristics of the pristine and irradiated (0.5 Mrad) specimens including transconductance $G_m$ (dashed lines) and drain current $I_{DS}$ (solid lines) versus drain voltage ($V_{DS}$) at a gate voltage of 2 V. }
	  \label{fig:elect}
\end{figure}

As a secondary means to confirm the evolution of strain and stress fields in the two devices, transconductance ($G_m$) and drain current ($I_{DS}$) measurements were performed on the two specimens. Fig.~\ref{fig:elect} shows these values versus drain voltage ($V_{DS}$) with application of a 2 V gate voltage. In the figure, we can see that transconductance and drain saturation current increase because of 0.5 Mrad $\gamma$-irradiation. As transconductance is directly proportional to the mobility of charge carriers, these results indicate a strain modification along the device channel. Lee et al.  found similar improvement below $25 \times 10^3$ rad with an increase in minority carrier lifetime and carrier diffusion length \cite{lee2017low}. 

\section{Summary}

Here we presented the first 2D field measurements of residual strain and stress concentrations within HEMTs using differential aperture X-ray microscopy. Residual lattice strains were measured in both pristine and irradiated specimens in order to evaluate the effects of $\gamma$-irradiation on residual stress concentrations. Developing these measurement capabilities is critical for optimizing the electronic and thermomechanical properties of microelectronic devices, in addition to determining the effects of residual stress on performance in environments in which radiation damage may occur. The particular findings in this work include:
\begin{itemize}
    \item The largest strain magnitude concentration in the GaN layer of the pristine HEMT device was underneath the transistor drain, while it was underneath the transistor gate in the irradiated specimen.
    \item Irradiation significantly decreased the measured lattice strains normal to the device surface in the GaN layer, likely due to radiation damage in the adhered gold electrode layers.
    \item Estimated residual stresses were of reasonable order of magnitude (10$^3$ MPa), providing confidence in the measurement technique.
\end{itemize}

\section*{Acknowledgments}

Aman Haque acknowledges partial support from the (a) Defense Threat Reduction Agency (DTRA) as part of the Interaction of Ionizing Radiation with Matter University Research Alliance (IIRM-URA) under Contract No. HDTRA1-20-2-0002 and (b) the U.S. National Science Foundation (ECCS No. 2015795). The content of the information does not necessarily reflect the position or the policy of the federal government, and no official endorsement should be inferred. Use of the Advanced Photon Source was supported by the U. S. Department of Energy, Office of Science, Office of Basic Energy Sciences, under Contract No. DE-AC02-06CH11357.

\section*{Data Availability}

The data that support the findings of this study are available from the corresponding author upon reasonable request.

\newpage
\clearpage

\bibliographystyle{elsarticle-num}
\bibliography{References.bib}

\end{document}